\documentclass[12pt]{iopart}
\usepackage{graphicx}

\begin{document}

\title{The Herbertsmithite Hamiltonian: $\mu$SR measurements on single
crystals}
\author{Oren Ofer$^1$, Amit Keren$^2$, Jess H. Brewer$^3$, Tianheng H. Han$%
^4 $ and Young S. Lee$^4$}

\begin{abstract}
We present transverse field muon spin rotation/relaxation measurements on
single crystals of the spin-1/2 kagome antiferromagnet Herbertsmithite. We
find that the spins are more easily polarized when the field is
perpendicular to the kagome plane. We demonstrate that the difference in
magnetization between the different directions cannot be accounted for by
Dzyaloshinksii-Moriya type interactions alone, and that anisotropic axial
interaction is present.
\end{abstract}

\address{$^1$ TRIUMF, 4004 Wesbrook Mall, Vancouver, BC V6T2A3, Canada} %
\ead{oren@triumf.ca} 
\address{$^2$ Department of Physics, Technion, Haifa
32000, Israel} 
\address{$^3$ Department of Physics and Astronomy, University of British Columbia,
Vancouver, V6T1Z1 Canada } 
\address{$^4$ Department of Physics, Massachusetts Institute of Technology,
Cambridge, Massachusetts 02139, USA}
\maketitle

After many years of searching for a good model compound for the spin-$1/2$
antiferromagnetic kagome magnet, it seems that the community is converging
on Herbertsmithite, ZnCu$_{3}$(OH)$_{6}$Cl$_{2}$, as the system closest to
ideal. Recently, a major scepticism was removed when it was shown that the
Zn ions in single crystals do not reside in the kagome plane \cite{Freedman}%
. However, it is not yet clear what exactly is the Hamiltonian controlling
the behavior of this system. Are the interactions isotropic or not? Is the
Dzyaloshinskii Moriya (DM) interaction \cite{marcos} relevant?

To address these questions the research must advance to single crystals.
These are available, but their size is still small, limiting the experiments
available for them. The high transverse field muon spin rotation ($\mu $SR)
technique is capable of overcoming this size problem, since the high field
helps focus the muon beam onto the small crystals. Here we report such
measurements. We find that the magnetic response of Herbertsmithite is very
anisotropic. We then analyze the magnetization data and show that a
non-isotropic diagonal interaction must be present in ZnCu$_{3}$(OH)$_{6}$Cl$%
_{2}$.

\begin{figure}[tb]
\begin{center}
\includegraphics{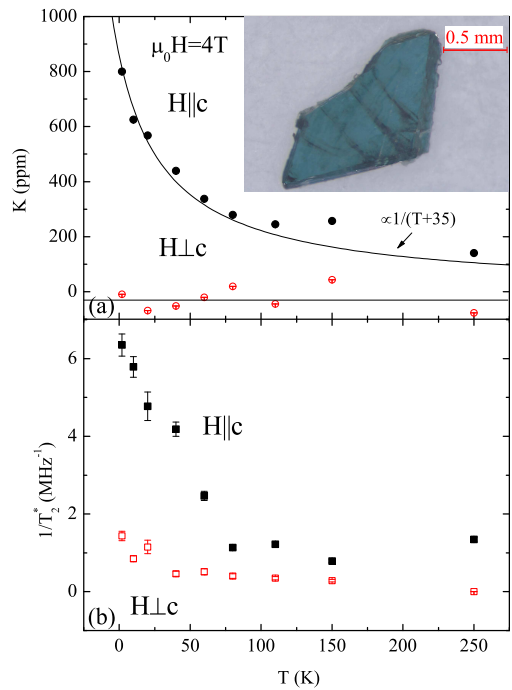}
\end{center}
\caption{(Color online) The temperature dependence of (a) the muon frequency
shift $K$ (errors are smaller than the symbol) and (b) relaxation rate $%
1/T_{2}^{\ast }$, for two orientations of the crystals (filled symbols for $%
\vec{\mathbf{H}}\parallel \widehat{\mathbf{c}}$, hollow for $\vec{\mathbf{H}}%
\perp \widehat{\mathbf{c}}$). The solid line represent a Curie-Weiss type
law. The inset shows a typical crystal.}
\label{fig:shift}
\end{figure}

Pioneering measurements on Herbertsmithite indicated a Curie-Weiss (CW)
temperature $\theta =-314$~K and a broad deviation from the high-temperature
CW behavior starting at $T\approx 75$~K. The nearest-neighbor super-exchange
interaction leads to a coupling of $J\approx 190$~K \cite{HeltonPRL}.
Extensive measurements on powder samples have found no evidence for
long-range magnetic ordering or spin freezing down to $20$~mK \cite%
{mendelsPRL,fabrice}, or a gap to excitations \cite{meCondMat,olariu}. A
recent Raman spectroscopic study on single crystals gives further evidence
for a gapless spin liquid state \cite{Wulferding}. Analysis of electron spin
resonance spectra using DM interactions \emph{only\/} suggests a sizable DM
vector of $D_{z}=15$~K \cite{ZorkoPRL}. Nuclear magnetic resonance spectra,
analyzed again using DM interactions \emph{only}, but with the addition of
defects arising from site-exchange, claimed $11\leq D\leq 19$~K \cite{Mila}.
However, magnetization measurements on oriented powders discovered a
dramatic difference in the magnetization between different directions \cite%
{mePRB} and demonstrated that DM is not the only perturbation to the
Heisenberg Hamiltonian. The measurements performed here on single crystals
are a clear improvement on the experiments with oriented powders.

\begin{figure}[tb]
\begin{center}
\includegraphics{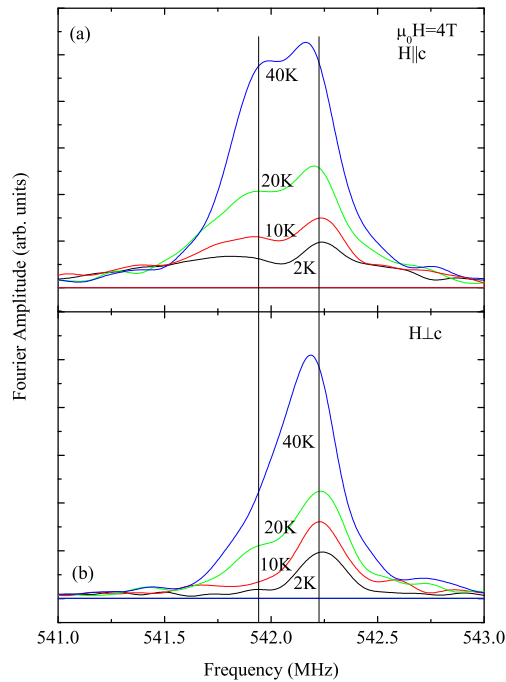}
\end{center}
\caption{(Color online) The temperature dependence of the Fourier transform
of the $\protect\mu $SR asymmetry data in a field of 4~T. (a) The spectra
obtained when the field $\vec{\mathbf{H}}$ is parallel to the $\widehat{%
\mathbf{c}}$-axis, which is normal to the mosaic (and thus to the kagome
plane) (b) The spectra obtained when the field $\vec{\mathbf{H}}$ is
perpendicular to the $\widehat{\mathbf{c}}$-axis.}
\label{fig:ft}
\end{figure}

The single crystals were measured by X-ray diffraction using a Bruker D8 AXS
single crystal diffractometer in order to reveal their crystallographic
axes. Subsequently a mosaic of 6 single crystals was created. In the mosaic,
the $\widehat{\mathbf{c}}$ axis is off by few degrees between crystals, and
we did not manage to keep a particular orientation of the a-b plane. One
face of a crystals is shown in the inset of Fig.~\ref{fig:shift}. In
general, all faces had similar size and the crystals shape is closer to a
cube rather than a slab or a needle. The dark lines seen on the crystal are
likely cracks or uneven steps, however x-ray refinement do find a single
phase. The $\widehat{\mathbf{c}}$ axis of the crystal is the direction
perpendicular to the kagome plane. 

The mosaic, held by a thin Mylar tape, was placed onto a low-background
sample holder on the M15 surface muon channel at TRIUMF, Canada. Transverse
field (TF) $\mu $SR spectra, where the field is applied perpendicular to the
muon spin direction, were gathered in the $T$ range between $2$~K and $250$%
~K in a constant field of $\mu _{0}H=4$~T. Thereafter, the mosaic was
rotated by 90$^{\circ }$ to probe the second orientation of the crystals. A
TF-$\mu $SR experiment is a sensitive probe of the magnetization $M$ of the
specimen through the precession frequency of the muon spin. The frequency
shift $K^{\alpha \alpha }(T)$ for a field in the $\alpha $ direction is
proportional to $M^{\alpha }/H^{\alpha }$ defined here as $\chi ^{\alpha
\alpha }(T)$. However, it should be noted that the ratio of shifts in
different directions is not the same as the ratio of susceptibilities in
those directions, since the shift is also determined by the muon-spin to
electronic-spin coupling. This coupling has a significant dipolar character.

Figure \ref{fig:ft} depicts the Fourier transforms of the $\mu $SR data
obtained at $T\leq 40$~K. At the highest $T$ (not shown) a symmetric peak at
542.1~MHz is seen for both field orientations. Below 150~K the peak becomes
asymmetric (not shown). Below 40~K, two clear peaks show up in the $\vec{%
\mathbf{H}}\parallel \widehat{\mathbf{c}}$ measurement [panel (a)]. This
happens only at 20~K in the $\vec{\mathbf{H}}\perp \widehat{\mathbf{c}}$
measurement [panel (b)]. In both cases, the emerging lower intensity peak
appears below 542.1~MHz. As the temperature is lowered, the low intensity
peak in the $\vec{\mathbf{H}}\parallel \widehat{\mathbf{c}}$ spectrum shifts
to even lower frequencies and broadens. In contrast, the low intensity peak
for $\vec{\mathbf{H}}\perp \widehat{\mathbf{c}}$ smears out quickly, and is
unseen at 2~K. The high intensity peak does not shift in either case. The
solid vertical lines help assess the shift. The low frequency peaks can be
assigned to muons that are influenced by the magnetic kagome planes, since
such a wipe out of the signal is typical of slowing down of spin
fluctuations, which is expected as $T$ decreases.

\begin{figure}[tb]
\begin{center}
\includegraphics{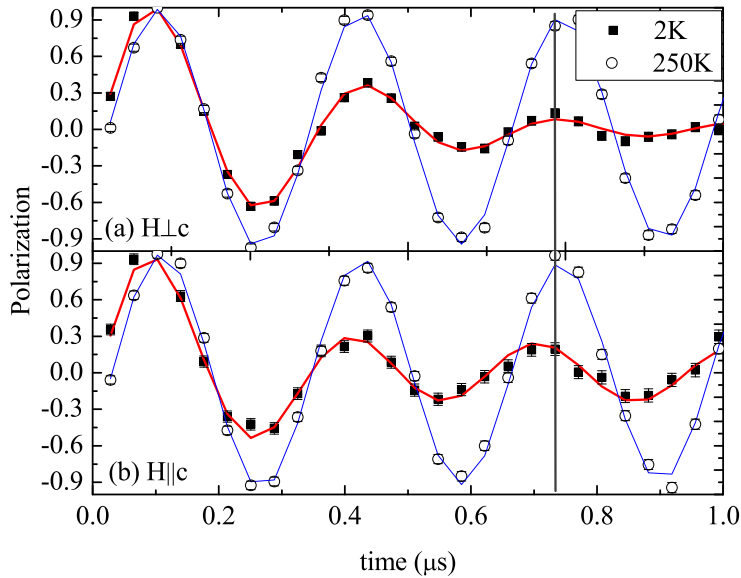}
\end{center}
\caption{(Color online) $\protect\mu $SR data in the time domain using a
rotating reference frame of 539 MHz, for the two different field directions
and two different temperatures: 250 and 2 K. (a) $\vec{\mathbf{H}}\perp 
\widehat{\mathbf{c}}$; no shift is detected between the two temperatures,
but an increase in relaxation is observed. (b) $\vec{\mathbf{H}}\parallel 
\widehat{\mathbf{c}}$; a shift in the frequency and increase in relaxation
are observed. The solid lines are fits of Eq. \protect\ref{eqfit} to the
data. The solid vertical line show when a shift is present and when it is
not.}
\label{fig:rawtdata}
\end{figure}

Raw data in the time domain for the $T=2\;$K and $T=250\;$K are shown in
Fig. \ref{fig:rawtdata}. These data are presented in a rotating reference
frame of 539 MHz. The vertical line indicates that in the $\vec{\mathbf{H}}%
\perp \widehat{\mathbf{c}}$ case [panel (a)] no change in the rotation
frequency is detected and only an increase in the relaxation is observed. In
the $\vec{\mathbf{H}}\parallel \widehat{\mathbf{c}}$ case [panel (b)] the
frequency does shifts and the relaxation increases. The reason for the
apparent increase in the frequency in Fig.~\ref{fig:rawtdata}b is that in
the time domain one sees the mean frequency. In the $\vec{\mathbf{H}}%
\parallel \widehat{\mathbf{c}}$ case the mean frequency shifts upward since
the low frequency peak diminishes faster than the high frequency peak. This
is not the case for $\vec{\mathbf{H}}\perp \widehat{\mathbf{c}}$. In light
of the Fourier transform the function 
\begin{eqnarray}
A_{\mathrm{TF}}(t) &=&A_{\parallel ,\perp }\exp (-\sqrt{t/T_{2\parallel
,\perp }^{\ast }})\cos (\omega _{\parallel ,\perp }t+\varphi _{\parallel
,\perp })  \nonumber \\
&&+A_{2}\exp (-(\sigma t)^{2}/2)\cos (\omega _{2}t+\varphi _{2})
\label{eqfit}
\end{eqnarray}%
is fitted to these data globally. The fit is also shown in Fig.~\ref%
{fig:rawtdata} using solid lines. The two terms represent an oscillating
signal relaxing as a root exponential, originating from the kagome planes,
and a Gaussian-relaxing signal stemming from a paramagnetic site. $A$, $%
T_{2}^{\ast }$, $\sigma $ and $\omega $ are the corresponding Asymmetry,
relaxation times, relaxation rate and frequencies for each peak in each
geometry, respectively. The total asymmetry is a common parameter in the
global fit. The ratio $A_{\parallel ,\perp }/A_{2}=3(1)$ indicates that most
muons sense the kagome planes. The very fast relaxation at early times makes
the data in the time domain look as if some asymmetry is lost. This
translates to a reduced Fourier amplitude in Fig.~\ref{fig:ft} upon cooling.
A fit in the time domain with a fixed $A_{\mathrm{TF}}(0)$ for all
temperatures overcomes this problem.

In Fig.~\ref{fig:shift}(a) we plot the frequency shift, $K_{\parallel ,\perp
}=(\omega _{2}-\omega _{\parallel ,\perp })/\omega _{2}$, versus
temperature. The shifts $K_{\parallel }$ and $K_{\perp }$ behave very
differently with decreasing temperature. $K_{\parallel }$ increases rapidly
with decreasing $T$ below 100~K and reaches 800~ppm at 2~K. In contrast, $%
K_{\perp }$ fluctuates and is very small. The behavior of $K_{\perp }$ is
not understood at the moment; however, the temperature average $\overline{%
K_{\perp }(T)}=-30(50)$~ppm hints that these might be muon site
fluctuations. $K_{\parallel }$ is an order of magnitude larger than the
macroscopic suscetibility $\chi $ at the same applied field as measured by a
SQUID \cite{Harry}. The fact that $K_{\parallel }\sim 4\pi \chi $ is, 
\textit{a priori}, a cause for concern. It might indicate that the muons are
working as a magnetometer and that the difference between $K_{\parallel }$
and $K_{\perp }$ is due to differences in the sample's demagnetization
factor $D$ in different directions. However, the two lines, clearly
resolved at 20~K for each direction, indicates that the muons in the site
with a shift is sensing the local susceptibility. If the muons were only
experiencing $D\chi $, we would not see different behavior at different
sites. Moreover, we do not expect differences in $D$ for different
directions due to the crystals shape. Thus, the behavior of $K_{\perp }$
and $K_{\parallel }$ indicates a very small spin response when the field is
in the kagome plane compared to a field perpendicular to this plane.

Since the temperature dependence of the shift is proportional to the local
susceptibility these results indicates that Herbertsmithite has an easy
axis. We fit $K_{\parallel }(T)$ to a CW type law and find $\theta _{cw}=-35$%
\ K. We also fit our $K_{\parallel }(T)$ data to a CW law with an offset. We
examined a free offset and an offset set by the $\overline{K_{\perp }}$
data. We found that $\theta _{cw}$ either does not change or becomes more
negative. We will continue our discussion using the minimal $\left\vert
\theta _{cw}\right\vert $ since it is sufficient for the conclusions of this
paper. This $\theta _{cw}$ is different from the one obtained by a SQUID on
powders. However, in Herbertsmithite different probes gave different
behavior \cite{meCondMat,olariu,mePRB,imai,ZorkoPRL} upon cooling so having
a new CW is not so surprising. 

Finally, in Fig.~\ref{fig:shift}(b) we present the relaxation rate $%
1/T_{2\parallel ,\perp }^{\ast }$. The relaxation in both directions is flat
and small down to 70\ K and then it increases. However, at low $T$, the
relaxation is very different for the two directions. Since $1/T_{2}^{\ast }$
is also proportional to the susceptibility \cite{generalMuSR} the relaxation
measurement indicate, again, that the system is anisotropic.

In order to understand this behavior we turn to an anisotropic Heisenberg
Hamiltonian with a DM interaction which we write as 
\begin{equation}
\mathcal{H}=\sum_{i}g\mu _{B}\vec{\mathbf{S}}_{i}\cdot \vec{\mathbf{H}}%
-\sum_{j\neq i}J_{\perp }\vec{\mathbf{S}}_{i}^{\perp }\cdot \vec{\mathbf{S}}%
_{j}^{\perp }+J_{z}S_{i}^{z}S_{j}^{z}+\mathbf{\vec{D}}_{ij}\cdot (\vec{%
\mathbf{S}}_{j}\times \vec{\mathbf{S}}_{i})\;,  \label{Hamil}
\end{equation}%
where $\vec{\mathbf{D}}_{ij}$ are the DM vectors and the second sum is over
neighboring spins (not bonds). This Hamiltonian can be written as $\mathcal{H%
}=g\mu _{B}\sum_{i}\vec{\mathbf{S}}_{i}\cdot \vec{\mathbf{H}}_{i}^{\mathrm{%
eff}}$, where the effective field is 
\begin{equation}
\vec{\mathbf{H}}_{i}^{\mathrm{eff}}=\vec{\mathbf{H}}-\frac{1}{g\mu _{B}}%
\bigl(\sum_{j\neq i}{\tilde{\mathbf{J}}}\vec{\mathbf{S}}_{j}+\vec{\mathbf{D}}%
_{ij}\times \vec{\mathbf{S}}_{j}\bigr)  \label{Heff}
\end{equation}%
and the anisotropic diagonal coupling is given by 
\begin{equation}
\widetilde{\mathbf{J}}=\left( 
\begin{array}{ccc}
J_{\perp } & 0 & 0 \\ 
0 & J_{\perp } & 0 \\ 
0 & 0 & J_{z}%
\end{array}%
\right) \;.  \label{Janiso}
\end{equation}%
When expressing $\vec{\mathbf{H}}^{\mathrm{eff}}$ one has to be careful
about the convention of $\vec{\mathbf{D}}_{ij}$ \cite{marcos}. We continue
with the mean-field approximation ($\vec{\mathbf{S}}\rightarrow -\vec{%
\mathbf{M}}/g\mu _{B}$). In principle the kagome unit cell has three
different atoms and we should allow a different $\vec{\mathbf{M}}$ for each
site. However, we are interested in the high temperature limit where it is
reasonable to assume that the magnetization of all ions is the same. We will
check this assumption at the end of the calculation. Thus 
\begin{equation}
\vec{\mathbf{H}}_{i}^{eff}=\vec{\mathbf{H}}+\frac{Z}{(g\mu _{B})^{2}}\left( 
\widetilde{\mathbf{J}}\vec{\mathbf{M}}+\vec{\mathbf{D}}_{i}\times \vec{%
\mathbf{M}}\right) \;,
\end{equation}%
where $\vec{\mathbf{D}}_{i}=\frac{1}{Z}\sum_{j}\mathbf{\vec{D}}_{ij}$ and $Z$
is the number of near neighbors. The magnetization is given by a self
consistent solution of the equation $\vec{\mathbf{M}}_{i}=\frac{C}{T}\vec{%
\mathbf{H}}_{i}^{eff}$ where $C=(g\mu _{B})^{2}\;S(S+1)/(3k_{B})$ is the
Curie constant. The solution is given by 
\begin{eqnarray}
\chi _{i}^{\bot \bot } &=&C\frac{(T-\theta _{\perp })(T-\theta
_{z})+D_{i}^{\perp \prime 2}}{(T-\theta _{z})(T-\theta _{\perp })^{2}+T\vec{%
\mathbf{D}}_{i}^{\prime 2}-{\vec{\mathbf{D}}_{i}^{\perp \prime 2}}\theta
_{\perp }-D_{i}^{z\prime 2}\theta _{z}}\;,  \nonumber \\
\chi _{i}^{zz} &=&C\frac{(T-\theta _{\perp })^{2}+D_{i}^{z\prime 2}}{%
(T-\theta _{z})(T-\theta _{\perp })^{2}+T\vec{\mathbf{D}}_{i}^{\prime 2}-{%
\vec{\mathbf{D}}_{i}^{\perp \prime 2}}\theta _{\perp }-D_{i}^{z\prime
2}\theta _{z}}\;,  \label{finalchi}
\end{eqnarray}%
where 
\[
\vec{\mathbf{D}}^{z,\perp \prime }=\frac{CZ}{(g\mu _{B})^{2}}%
\vec{\mathbf{D}}^{z,\perp }\;. 
\]%
Here $\vec{\mathbf{D}}^{z,\perp }$ are the DM components in the $%
z $ direction and in the kagome plane, respectively.

It is simple to see that there are three different $\vec{\mathbf{D}}_{i}$ {in%
} the lattice. Therefore, $\chi _{i}^{zz}$ and $\chi _{i}^{\bot \bot }$ or $%
\vec{\mathbf{M}}_{i}$ must be site dependent, unless we are in the limit of
high temperatures, $(T-\theta )^{2}\gg (\mathbf{D}_{i}^{z,\perp \prime
})^{2} $ where 
\begin{equation}
\chi ^{zz}=\frac{C}{(T-\theta _{z})}\rm{~and~}\chi ^{\bot \bot }=\frac{C}{%
T-\theta _{\bot }}{.}  \label{SimpleChi}
\end{equation}%
In this case the DM interaction does not affect the CW temperatures which
are determined from the high $T$ data. In our experiment at $T=65$\ K, $%
T-\theta _{cw}=100$~K and $\mathbf{D}_{ij}$ is estimated to be on the order
of 10~K \cite{ZorkoPRL}. Therefore, $(T-\theta _{cw})^{2}$ is two order of
magnitude larger than $(\mathbf{D}_{i}^{z,\perp \prime })^{2}$, and the high
temperature limit is valid. Since the shifts, hence the susceptibilities in
the two directions are different at $T=65$\ K we conclude that $\theta
_{z}\neq \theta _{\perp }$, that $J$ must have axial anisotropy, and that DM
interaction cannot explain the anisotropy in the measurements. This does not
mean that the DM interaction, which is allowed by the symmetry of the kagome
lattice, is not present in addition to the axial anisotropy \cite{Maged}.

In conclusions, $\mu $SR measurements were performed on single crystals of
Herbertsmithite. We find no clear signature of a magnetic transition down to 
$T<10^{-2}J$~, as was found for powders (down to $10^{-4}J$). Anisotropic
spin susceptibility with an easy axis is revealed. By mean-field
approximations, we show that this phenomenon cannot be due to DM
interactions alone, and an anisotropic super-exchange $J$ is needed. Our
data calls for more work on larger crystals where one can avoid using a
mosaic of crystal and the anisotropy can be measured more accurately. Such
crystal are becoming available. It will also be useful to have theoretical
work on the high temperature behavior of the magnetization in different
directions, using Eq.\ \ref{Hamil} and more accurate methods than mean field.

\ack We are grateful to the staff of the TRIUMF CMMS Facility for assistance
with the $\mu ^{+}$SR experiments. We are also grateful to P. Mendels for
helpful discussion. AK and YSL would like to thank the Israel-US Binational
Science Foundation for funding this work. JHB is supported at UBC by NSERC
of Canada and (through TRIUMF) by NRC of Canada.


\end{document}